\documentclass[submission]{IEEEtran}
\ifCLASSINFOpdf
\else
  \usepackage{graphicx}

\fi
\usepackage{amssymb}
\usepackage[cmex10]{amsmath}
\begin{document}
%
\title{Finding overlapping communities in networks using evolutionary method}
%
%
%

\author{Weihua~Zhan,
        Huahui~Chen,
        Jihong Guan,
        and Guang Jin
\thanks{W. Zhan, H. Chen, and G. Jin is with the College
of Information Science and Engineering, Ningbo University, Ningbo,
315211, China e-mail: (zhanweihua@nbu.edu.cn).}
\thanks{J. Guan is with Tongji University, Shanghai.}
\thanks{Manuscript received September 19, 2012; revised January 11, XXXX.}}

%
%

\markboth{Journal of \LaTeX\ Class Files,~Vol.~6, No.~1, January~2012}%
{Shell \MakeLowercase{\textit{et al.}}: Bare Demo of IEEEtran.cls for Journals}
%



\maketitle

\begin{abstract}
Community structure is a typical property of many real-world networks, and has become a key to understand the dynamics of the networked systems. In these networks most nodes apparently lie in a community while there often exists a few nodes straddling several communities. An ideal algorithm for community detection is preferable which can identify the overlapping communities in such networks. To represent an overlapping division we develop a encoding schema composed of two segments, the first one represents a disjoint partition and the second one represents a extension of the partition that allows of multiple memberships. We give a measure for the informativeness of a node, and present an evolutionary method for detecting the overlapping communities in a network.
\end{abstract}


%
\IEEEpeerreviewmaketitle

\section{Introduction}

\IEEEPARstart{A}{s} a unified tool for study various complex systems, networks have attracted tremendous attentions during the last ten years~\cite{Albert02,Newman03,Doro03,Boccaletti06}, with nodes representing the units and edges denoting diverse
interactions between these units. In social networks edges often capture various social relation between individuals; In technology networks~(such as Internet) an edge may correspond to a physical connection~(or communication linkage) between two sites.

Community structure is an important topological feature of real networks, which refers to the natural clusters of nodes such that the connections within clusters are significantly more dense than the connections between clusters~\cite{Girvan02}. Due to the intimate affiliation with function decomposition and various dynamics of systems, community structure detection has been extensively study. A variety of methods for detecting communities have been
proposed based on different principle and heuristics, such as
divisive method based on betweenness~\cite{Girvan02}, methods based on
modularity optimization by simulated annealing~\cite{Guimear05}, spectral method~\cite{Newman06a,Newman06b} or
extremal optimization\cite{Danon05},
methods based on dynamical process including random walks~\cite{Zhou03} or
synchronization~\cite{Arenas06} and methods based on different
formal definitions of community~\cite{Radicchi04,Palla05}.

For most existing methods a common assumption is that community structure is a disjoint division of the network, that is, any node should only belong to a community. However, it may be not the case for many real networks. In
scientific collaboration network, for instance, an energetic scientist would have participated several research groups with different concerns. Hence an ideal algorithm for community detection should be able to automatically find an accurate overlapping division of the network if its community structure is indeed overlapping.

Given a quality function such as modularity for a division of network, community detection is frequently formulated as an optimization problem. Since it has been proved an NP-Hard problem, various heuristics have been employed, including simulated annealing, spectral method, and evolutionary method. As opposed to other heuristics, evolutionary methods have higher accuracy and stronger stability arising from their search mechanize based on population.

In this paper, We extend the evolutionary method for community detection, MAGA, to detect overlapping communities. In section 2, we review several measures for an overlapping partition of a network. In section 3, we describe our
evolutionary method for detecting overlapping communities in detail. In section 4, we test our method in model network and several real networks. At last, the conclusion is given.

\section{Overlapping Community Structure}
A network  $\mathcal{N} $ can be represented as a graph $G=(V,E)$, where $V$ is a set of nodes, and $E$ is a set of edges. Community Structure of the network corresponds to a division of the set of nodes E, such that for any group of nodes (i.e., community), the density of inner links is significantly higher than that of exterior links.
Formally, community structure contains $k$ communities can be read as
\begin{equation}\label{eq:DefCommunity}
\mathcal{C}=\{C_{1},C_{2},\ldots,C_{k}\}
\end{equation} such that
\begin{equation}
 V=C_{1}\bigcup C_{2}\bigcup \cdots C_{k},\\\forall i,1\leq i\leq k, C_{i}\neq\varnothing
\end{equation}
and
\begin{equation}\label{eq:DisjointCondition}
V_{i}\cap V_{j}=\varnothing,\forall i,j,1\leq i,j\leq k, i\neq j
\end{equation}
Equations.~(\ref{eq:DefCommunity})-(\ref{eq:DisjointCondition}) defines a disjoint division of $V$, and most existing methods for community detection can propose a such division. By removing the restriction from Equation~(\ref{eq:DisjointCondition}), {\it Overlapping community structure} allows for communities being overlapping.

\section{Related work}
\subsection{Evolutionary method for community detection}

\subsection{Overlapping communities detection}
To detect overlapping communities, Palla et al\cite{Palla05} presented a clique percolation method where a community is a union of some adjacent k-cliques~(complete subgraph with $k$ nodes) in the network, and thereby called as a k-clique community. This method has been extensively applied to the analysis of social networks and biology network. However, it has some limits: (1) The divisions found with different k values generally differ from each other, and then an incidental question is which is best; (2) For any k value there almost always exist some dissociative nodes that have no membership in any k-clique community. Similar to detecting disjoint community, the detection of overlapping community also can be formulated as an optimization problem given an appropriate measure for an overlapping division. Nicosia et al\cite{Nicosia09} extended the modularity to overlapping case, and then proposed a genetic algorithm to optimize their quality function. Shen et al\cite{Shen09} also presented an overlapping measure and then employed the Blondel's algorithm to optimize it. In addition, Zhang et al successively presented a fuzzy c-means method\cite{Zhangs07a} and negative matrix factorization method~\cite{Zhangs07b} for finding a good overlapping division. Recently, several methods based on extend a disjoint division of network to an overlapping division have been proposed.

\section{Measures for an overlapping division}

To measure the goodness of a partition of a network, modularity was proposed by Newman and Girvan~\cite{Newman041} which has been widely used as the objective function for
for community detection approaches based on optimization. There also exist other measures for the quality of a disjoint division, such as hamiltonian of potts model and absolute potts model, modular density~\cite{Lizp08}. Several definitions of overlapping modularity have been proposed for detecting overlapping communities. These measures for overlapping divisions are useful, but they may suffer the same resolution limit as modularity.

Modularity was proposed by Newman and Girvan~\cite{Newman041} to measure the goodness of a partition of a given network, which has been widely used as the objective function for
those methods for community detection based on optimization. The definition of modularity is based on the idea that the true community structure of the network should correspond to a statistically surprising arrangement of edges, that is, the number of actual links within
communities should be significantly beyond that of expected links of
a null model. Configuration model, an extensively used null model, is employed in the definition of modularity. Let $k_i$ be the degree of nodes \emph{i}, \emph{L} the
total number of edges, then in the null model the expectation of
edges presenting between nodes \emph{i} and \emph{j} is
$\frac{k_ik_j}{2L}$. The modularity thus can be written as follows:
\begin{equation}\label{eqmodularity}
Q=\frac{1}{2L}\sum_{C_i\in\mathcal{C}}\left(e_c^{in}-e_c^{exp}\right)
\end{equation}
where $e_c^{in}$ and $e_c^{exp}$ are the number of inner links in the subgraph \emph{c} and that of the expectation of inner links, which are counted as $\sum_{i,j\in G_c}A_{ij}$ and $\sum_{i,j\in G_c} \frac{k_ik_j}{2L}$, respectively. \emph{Q} is the sum of the difference over $|\mathcal{C}|$ groups of
the specific partition. The maximum value of \emph{Q} is 1, and a value
approaching 1 indicates strong community structure. Conversely,
when the number of within-community edges is no better than random case,
\emph{Q}=0, and a value approaching 0 implies weaker community structure
or indivisibility. For a network with strong community
structure, it normally falls in the range from around 0.3 to 0.7.

Since the above definition of modularity is actually designed for simple networks,
some variations have been presented for various types of network. To identify
overlapping community structure, it require redefine the number of inner links and the expectation of links
in a community. The number of inner links in community of $G_c$ can be counted as
\begin{equation}
e_c^{in}=\sum_{i,j\in G_c}S_{ic}S_{jc}A_{ij},
\end{equation}
and the expectation number of inner links reads
\begin{equation}
e_c^{exp}=\frac{\left(\sum_{i,j\in G_c}S_{ic}S_{jc}A_{ij}+\sum_{i\in G_c,j\not\in G_c}S_{ic}A_{ij}\right)^2}{2L}
\end{equation}

The main difference among these is the calculation of the expectation links.
\section{Method}

Once given an appropriate measure for the quality of a partition for a network, the problem of community detection can be cast as an optimization problem. Then the key to identify an accurate partition of a network is to find an effective optimization method.

The existing measures, such as modularity and hamiltonian, are nonlinear and the optimization on which is NP hard. Therefore, various heuristic methods have been employed on community detection. Generally, those heuristics with high time cost, including simulated annealing and tabu search, can get more accurate results while those fast algorithms would obtain lower accurate results.

It is notable that an evolutionary method, called MAGA~(Modified adaptive genetic algorithm), was recently presented, in contrast with simulated annealing method which shows a higher accuracy on the widen used set of real benchmark networks and can deal with a network with more large scales.
As opposed to the with state-of-the-art heuristics, Single
Step Multi Level algorithm~(SS-ML)~\cite{Noack09} and Label Propagation algorithm for unipartite networks (LPAm+)~\cite{Liu10b}, it appears more stable on the test set as standard genetic algorithms. In the following, we presented the MAGA*, which is the extent of the MAGA for approaching the detection of overlapping communities.

\subsection{Encoding schema}
We first consider the representation of an overlapping partition of a network. Let the number of communities be c in the partition, the size of the network~(i.e., the number of the nodes in the networks) be n.

A simple encoding schema is that a chromosome consists of $n\times{c}$ loci, each of which indicates the belonging coefficient of node $i$ to community $c$ \cite{Nicosia09}. This representation requires $c$ to be no less than the number of true communities of the network, and which would result in a high cost in both time and space. On the other hand, it needs an extra repairing operator after performing genetic operators as the latter may produce illegal individuals.

Pizzuti~\cite{Pizzuti09} presented an indirect representation of overlapping communities, wherein the network is translated into a line graph and a partition of the latter
corresponds to an overlapping partition of the former. The partition of the line graph employs  locus-based adjacency representation, which will prevent the production of illegal individuals.
In a line graph a node corresponds to an edge in the original network and a link between two nodes stand for two edges has a same ends in the original network. It follows that a node with degree k in the original network will produce a k-clique. Therefore, the line graph is larger than the original one in size and has much more links, which increases the complexity of the problem.

\begin{figure}[htbp]
\centering
\includegraphics[width=2.5in]{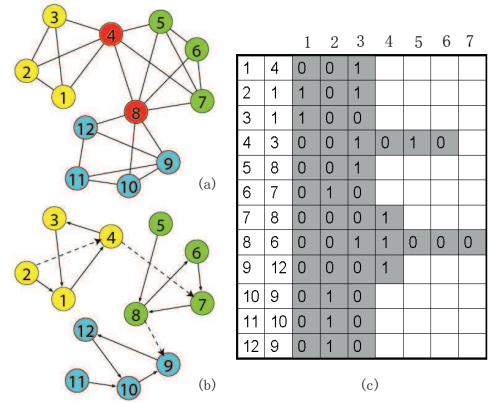}
\caption{Example network with overlapping community structure (a)Example network and its overlapping communities, $\pi^{*}=\{\{1,2,3,4\},\{4,5,6,7,8\},\{8,9,10,11,12\}\}$£¬
nodes 4 and 8 are overlapping nodes. (b) Sets of nodes connected by solid links form a disjoint partition, which contains three basic community, dashed lines 4--7 and 8--9
express overlapping relation. (c) Encoding schema. In the table the first column is the labels of nodes, and the rest of the table represent the chromosome representing the overlapping partition. The second column represents the disjoint partition, while the grey cells encode overlapping information which take values 0 or 1. The upper numbers indicate the k nodes of a node. The cell (i,j) indicates whether the nodes i is adhere to its j-2 neighbor.
} \label{fig:OverlapExample}
\end{figure}

Here, we present a coding schema that generalizes the locus-based adjacency representation to overlapping communities and has no need to the transformation from a network to its line graph.
This representation is based on the following idea:
\begin{itemize}
  \item A disjoint partition can be extracted from an overlapping partition;

  In an overlapping partition there are a few overlapping nodes while most of nodes are non-overlapping nodes. By assigning those overlapping nodes to only a single community, a disjoint partition can be obtained. As shown in figure~\ref{fig:OverlapExample}, the overlapping community structure of the network is $\pi^{*}=\{\{1,2,3,4\},\{4,5,6,7,8\},\{8,9,10,11,12\}\}$. One can from the overlapping partition obtain a disjoint one, ~$\pi=\{\{1,2,3,4\},\{5,6,7,8\},\{9,10,11,12\}\}$.
  \item Conversely, an overlapping partition can be educed from an disjoint partition.

  Take the network above as an example again. If the disjoint partition $\pi$ of the network has been obtained, then the overlapping community structure can be obtained by assign the nodes 4 and 8 to more than one communities. Actually, some recent work on identifying overlapping communities is essentially based on the presumption that a good overlapping partition of a network can be extended from a good disjoint one by determining a few overlapping nodes\cite{Gregory09,Wei09,Wang09b}.
\end{itemize}

For convenience, a \textbf{primary partition}, $\pi$, refers to a disjoint partition and an \textbf{extent partition} of $\pi$ is an overlapping partition extended from the primary partition $\pi$, denoted by $\pi^{*}$. To represent an overlapping partition, a chromosome consists of two segments. The first segment represents a primary partition, denoted by \textbf{P}. By locus-based adjacency representation, it consists of $n$ loci, each of which corresponds to a node and the allele indicates the neighbor node to which the node adheres. The second segment, denoted by \textbf{O},  represents the overlapping information between multiple communities. Similar to locus-based adjacency representation, node $i$ adhering to node $j$ implies that nodes i is a member of the community in the primary partition node j belong to. For a node with degree $k$, it owns k loci in the segment whose alleles take values of 0 or 1 indicates whether the node adheres to the corresponding neighbor.

As shown in figure \ref{fig:OverlapExample}~(b), in addition to node 3 in the same primary community node 4 adheres to node 7 outside the primary one. In this way, the primary community $\{5,6,7,8\}$ becomes an extend community $\{4,5,6,7,8\}$. Similarly, node 8 adheres to node 9 outside its primary community, which makes the primary community $\{9,10,11,12\}$ being an extended community $\{8,9,10,11,12\}$.

\subsection{Informativeness measures for nodes}
A practical problem for applying genetic algorithms is how to set the parameters since it can significantly affect the performance. To deal with this problem, Szeto and Zhang \cite{Szeto06} presented an adaptive genetic algorithm, MOGA~(Mutuation only Genetic Algorithm). Later, Law and Szeto~\cite{Law08} presented for the MOGA a framework that includes crossover operation. In both these two evolution algorithms probability of mutation on loci in a chromosome is not uniform, but varied with the informativeness of each locus. An allele standard deviation was proposed to measure the informativeness. Although this measure can work well in optimization of continuous variable, it would often result in a problem relating to the allele values for discrete optimization. Instead, in the MAGA the informativeness of a locus is measure by the bias between the actual distribution of alleles on current population and the random distribution, namely the Kullback-Leiber divergence between these two distributions. Let the distribution of alleles at locus \emph{i} be $\mathcal{P}_i$, the random one be $\mathcal{Q}_i$, then the informativeness of locus \emph{i} is
\begin{equation}
\mu_{i}=KLD(\mathcal{P}_i||\mathcal{Q}_i)=\sum_{x}P(X_i=x)\log |X_i|\cdot P(X_i=x)
\end{equation}
When applied to community detection, a node in a network corresponds to a locus, whose allele indicates which neighbor node it adheres to.

For easily describing the genetic algorithm proposed later, we introduce two measures for the informativeness about nodes. \emph{Primary informativeness}~(PINF) of node \emph{i} refers to the Kullback-Leibler divergence of the locus in the segment \textbf{P}, denoted by $u_{i}^P$. Then, it has $u_{i}^P=\mu_{P_i}$. In contrast, the definition of \emph{overall information}~(OINF) of a node appears complicated, which reflects the overall information of the node encoded by the two segments of chromosomes. This measure should consider two factors: the number of memberships, and the neighbor node in a extended community it adheres to. The first factor can be quantified by the Kullback-Leiber divergence of membership number of node \emph{i},
\begin{equation}
KLD_{M_i}=\sum_{k\le K}\widehat{P}(M_{i}=k)\log K\cdot\widehat{P}(M_{i}=k).
\end{equation}
where $K$ is the allowed maximum of memberships of a node in the network. Let the average number of memberships of the node \emph{i} be $\overline{M_i}$. We can define for the communities that node \emph{i} belongs to $M_i$ bias, $u_{i,1}, u_{i,2},\ldots,u_{i,M_i}$ as
\begin{equation}
u_{i,k}=\frac{\sum_{j\in Nbs(i)}P(x=j)\log \overline{h_{i,k}}\cdot P(x=j)}{\log \overline{h_{i,k}}}
\end{equation}
where $\overline{h_{i,k}}$ is the average number of alleles allowed, defined by
\begin{equation}
\setlength{\nulldelimiterspace}{0pt}
h_{i,k}=\left\{\begin{IEEEeqnarraybox}[\relax][c]{l's}
|Nbs(i)|, &for $k=1$,\\
\sum_{r}\left(|Nbs(i)|-\sum_{j=1}^{k-1}|Nbs{i,j}|\right), &for $k>1$%
\end{IEEEeqnarraybox}\right.
\end{equation}
Then, the overall information can be defined as
\begin{equation}
u_{i}^{o}=\frac{\sum_{k\le M_i}u_{i,k}}{M_i}
\end{equation}
\subsection{Mutation, and reassignment operator}

\section{Results}

\section{Conclusion}


%

\appendices
\section{Proof of the First Zonklar Equation}
Appendix one text goes here.

\section{}
Appendix two text goes here.

\section*{Acknowledgment}

The authors would like to thank...

\ifCLASSOPTIONcaptionsoff
  \newpage
\fi



%

%

\begin{IEEEbiography}{Weihua Zhan}
Biography text here.
\end{IEEEbiography}

\begin{IEEEbiographynophoto}{John Doe}
Biography text here.
\end{IEEEbiographynophoto}


\begin{IEEEbiographynophoto}{Jane Doe}
Biography text here.
\end{IEEEbiographynophoto}




\end{document}